\documentclass[iop]{emulateapj}
\usepackage{natbib}
\usepackage{color}
\usepackage[english]{babel}
\usepackage[normalem]{ulem}
\usepackage{blindtext}
\usepackage{textgreek}
\newcommand{\etal}{et\,al.}
\newcommand{\halpha}{H$\alpha$}
 
\newcommand{\lsim}{\raise0.3ex\hbox{$<$}\kern-0.75em{\lower0.65ex\hbox{$\sim$}}}
\newcommand{\gsim}{\raise0.3ex\hbox{$>$}\kern-0.75em{\lower0.65ex\hbox{$\sim$}}}
\newcommand{\msun}{M$_{\odot}$}

\newcommand{\kms}{km\,s$^{-1}$}
\begin{document}
\slugcomment{Accepted for publication in the Astrophysical Journal Letters}
%-----------------------------------------------------------------------------%
\title{First Characterization of the Neutral ISM in Two Local Volume 
Dwarf Galaxies}
%-----------------------------------------------------------------------------%

%-----------
\author{Lilly Bralts-Kelly, Alyssa M. Bulatek, Sarah Chinski, Robert N.
  Ford, Hannah E. Gilbonio, Greta Helmel, Riley McGlasson, Andrew Mizener,
  John M. Cannon}
\affil{Department of Physics \& Astronomy, Macalester College, 
1600 Grand Avenue, Saint Paul, MN 55105, USA}
\email{jcannon@macalester.edu}

\author{Serafim Kaisin, Igor Karachentsev}
\affil{Special Astrophysical Observatory of RAS, Nizhnij Arkhyz, KChR, 369167, Russia}
\email{ikar@sao.ru, skai@sao.ru}

\author{Grant Denn}
\affil{Department of Physics, Metropolitan State University of Denver, 
P.O. Box 173362, Denver, CO 80217, USA}
\email{gdenn@msudenver.edu}

%-----------------------------------------------------------------------------%
\begin{abstract}
%-----------------------------------------------------------------------------%

We present the first HI spectral line images of the nearby,
star-forming dwarf galaxies UGC\,11411 and UGC\,8245, acquired as part
of the ``Observing for University Classes'' program with the {Karl
  G. Jansky Very Large Array (VLA}\footnote{The National Radio
  Astronomy Observatory is a facility of the National Science
  Foundation operated under cooperative agreement by Associated
  Universities, Inc.}).  These low-resolution images localize the HI
gas and reveal the bulk kinematics of each system.  Comparing with HST
broadband and ground-based H$\alpha$ imaging, we find that the ongoing
star formation in each galaxy is associated with the highest HI mass
surface density regions.  UGC\,8245 has a much lower current star
formation rate than UGC\,11411, which harbors very high surface
brightness H$\alpha$ emission in the inner disk and diffuse, lower
surface brightness nebular gas that extends well beyond the stellar
disk as traced by HST. We measure the dynamical masses of each galaxy
and find that the halo of UGC\,11411 is more than an order of
magnitude more massive than the halo of UGC\,8245, even though the HI
and stellar masses of the sources are similar.  We show that UGC\,8245
shares similar physical properties with other well-studied low-mass
galaxies, while UGC\,11411 is more highly dark matter dominated.  Both
systems have negative peculiar velocities that are associated with a
coherent flow of nearby galaxies at high supergalactic latitude.

\end{abstract}						


%-----------------------------------------------------------------------------%
\section{Introduction}
\label{S1}
%-----------------------------------------------------------------------------%

Nearby dwarf galaxies provide unique insights into the physical
properties of low-mass halos.  Those systems which can be resolved
into individual stars allow the complex relationships between active
star formation and the surrounding interstellar medium (ISM) to be
examined in detail.  Similarly, in those systems where the bulk rotation 
velocity can be extracted, the distributions of dark and baryonic matter
can be examined on a spatially resolved basis.  

Major single-dish HI surveys have now cataloged thousands of gas-rich
objects in the local volume (e.g., HIPASS -- {Meyer
  \etal\ 2004}\nocite{meyer04}; ALFALFA -- {Giovanelli
  \etal\ 2005}\nocite{giovanelli05}).  From the spatially resolved
perspective, multiple recent and ongoing HI surveys have studied the
physical properties of many low-mass galaxies in the local volume
\citep[e.g.,][]{swaters2002, begum2008, cannon11, ott2012, hunter2012,
  kirby2012, lelli14, wang17}.  However, each HI survey has its own
selection criteria, and therefore not all local dwarf galaxies have
been examined in the HI spectral line on a spatially resolved
basis. Some star-forming, gas-rich dwarf galaxies possess extensive
observations at various wavelengths but have yet to have their neutral
ISM examined in detail.  In this manuscript, we present the first
observations of the HI 21\,cm emission line in two such galaxies:
UGC\,11411 and UGC\,8245 (see Table~\ref{table1} for a summary of
physical properties).  Note that each galaxy has a very low
(barycentric) recessional velocity; the proximity of Milky Way
foreground HI gas may have precluded detailed HI studies to date.

These galaxies were selected for study based on simple selection
criteria: each hosts ongoing star formation and has a direct distance
measurement from resolved Hubble Space Telescope (HST) photometry that
places the galaxy within the local volume.  Further, to date, no
measurements (single-dish or interferometric) of their HI gas contents
have been published.  These systems thus increase the number of nearby
dwarf galaxies with detailed observations of the neutral gas
component.

%-----------------------------------------------------------------------------%
\section{Observations and Data Handling}
\label{S2}
%-----------------------------------------------------------------------------%

HI spectral-line observations of UGC\,11411 and UGC\,8245 were
acquired with the VLA in the D configuration in February and March of
2017 for programs TDEM0023 and TDEM0024, respectively.  The data sets
are part of the ``Observing for University Classes'' program, a
service provided by NRAO as an opportunity for courses teaching radio
astronomy theory to acquire new observations to be analyzed by
students \citep{cannon17}.

Data handling and imaging followed the prescriptions described in
\citet{mcnichols16} and \citet{teich16}.  Briefly, a 4.0 MHz bandwidth
is separated into 1024 channels, delivering a native spectral
resolution of 0.86 km\,s$^{-1}$\,ch$^{-1}$.  The primary calibrator
was 3C\,286; the phase calibrators were J1748$+$7005 and J1323$+$7942
for UGC\,11411 and UGC\,8245, respectively.  The total on-source
integration time was $\sim$1.5 hours for each source.  The VLA data
were calibrated using standard prescriptions in the
CASA\footnote{https://casa.nrao.edu} environment.  Continuum
subtraction was performed in the $uv$ plane using a first-order fit to
line-free channels bracketing the galaxy in the central 50\% of the
bandpass.

Each dataset was imaged via the CASA task CLEAN with 10\arcsec\ pixels
at 2 \kms\ velocity resolution.  Cleaning was performed to 2.5 times
the rms noise per channel in line-free channels.  The 
synthesized beam sizes and rms noises are
61.68\arcsec\,$\times$\,44.30\arcsec\ and 3.3 mJy\,Bm$^{-1}$
(UGC\,11411) and 60.58\arcsec\,$\times$\,40.04\arcsec\ and 3.0
mJy\,Bm$^{-1}$ (UGC\,8245).  The resulting cubes were threshold
blanked at the 2.5\,$\sigma$ level and then further blanked by hand to
isolate emission that is coherent in velocity space.  Finally, the
cubes were collapsed to create traditional moment maps representing
HI mass surface density and intensity weighted velocity field.

Both target galaxies have low recessional velocities (see
Table~\ref{table1} and further discussion in \S~\ref{S4}).  We thus
pay particular attention to the possibility of contamination of each
line profile by Milky Way foreground HI gas.  For UGC\,11411 the
emission from the galaxy is cleanly separated in velocity space from
Milky Way gas; the lowest velocity HI gas detected from the source is
located at $+$35 \kms.  For UGC\,8245, however, there is moderate
contamination by Milky Way gas; specifically, in the velocity range of
$-$10 \kms\ to $+$6 \kms, there is extended foreground gas detected
over much larger angular scales than the galaxy.  At these velocities
we are able to spatially differentiate this extended gas from the
comparatively compact emission from the target galaxy.  We note that
at these velocities, the absolute flux from UGC\,8245 is less certain
than at the more negative velocities where the galaxy is cleanly
separated from the Milky Way foreground.  Based on the symmetry of the
HI gas about the systemic velocity (see Section \S~\ref{S3}), the
contamination appears to be a minor effect.

We compare our new HI imaging with archival ground-based and HST
optical imaging.  The optical imaging was acquired with the SAO 6.0\,m
telescope; detailed descriptions of data handling can be found in
{Karachentsev \& Kaisina (2013)}\nocite{KK13}.  The flux-calibrated
and continuum-subtracted \halpha\ images were registered to the WCS
coordinate system of the HST images to create the images presented in
this work.

\begin{deluxetable}{lcc}  
\tablecaption{Basic Characteristics of UGC\,11411 and UGC\,8245} 
\tablewidth{0pt}  
\tablehead{ 
\colhead{Parameter} &\colhead{UGC\,11411} &\colhead{UGC\,8245}}    
\startdata      
Right ascension (J2000)          &19$^{\rm h}$ 08$^{\rm m}$ 42.$^{\rm s}$3  &13$^{\rm h}$ 08$^{\rm m}$ 36.$^{\rm s}$2\\
Declination (J2000)              &+70\arcdeg 17\arcmin 02\arcsec &+78\arcdeg 56\arcmin 14\arcsec \\    
Distance (Mpc)                   &6.58\,$\pm$\,0.12\tablenotemark{a} &4.72\,$\pm$\,0.07\tablenotemark{a} \\
V$_{\rm HI}$ (\kms)\tablenotemark{b}              &88\,$\pm$3 &$-$26\,$\pm$\,3\\ 
M$_{\rm B}$ (Mag.)                &$-$14.01\tablenotemark{c} &$-$13.76\tablenotemark{c}\\
M$_{\star}$ (\msun)               &7.1\,$\times$\,10$^{7}$\tablenotemark{c} &7.3\,$\times$\,10$^{7}$\tablenotemark{d}\\
S$_{\rm HI}$ (Jy km\,s$^{-1}$)     &3.0\,$\pm$\,0.3 &2.6\,$\pm$\,0.3\\
M$_{\rm HI}$ (\msun)              &(3.1$\pm$\,0.46)\,$\times$\,10$^7$ &(1.4$\pm$\,0.21)\,$\times$\,10$^7$\\
M$_{\rm dyn}$ (\msun)\tablenotemark{e}     &2.3\,$\times$\,10$^{9}$ &1.7\,$\times$\,10$^{8}$\\
M$_{\rm dyn}$ (\msun)\tablenotemark{f}     &1.4\,$\times$\,10$^{9}$ &1.5\,$\times$\,10$^{8}$
\enddata     
\label{table1}
\begin{small}
\tablenotetext{a}{\citet{tully13}} 
\tablenotetext{b}{Systemic velocity based on the new HI images presented in this work.}
\tablenotetext{c}{\citet{karachentsev13}}
\tablenotetext{d}{Revising the stellar mass from \citet{cook14} for
  the adopted distance of 4.72\,$\pm$\,0.07 Mpc.  This value can be
  compared to M$_{\star}$ $=$ 6.0\,$\times$\,10$^{7}$ \msun\ when applying the
  same techniques as used to determine the stellar mass of UGC\,11411
  \citep{karachentsev13}.}
\tablenotetext{e}{Derived using the HST-based 
  inclinations of $i$ $=$ 37\arcdeg\ and $i$ $=$ 75\arcdeg\ for
  UGC\,11411 and UGC\,8245, respectively (see \S~\ref{S3.2}).}
\tablenotetext{f}{Derived using the ground-based 
  inclinations of $i$ $=$ 52\arcdeg\ and $i$ $=$ 90\arcdeg\ for
  UGC\,11411 and UGC\,8245, respectively (see \S~\ref{S3.2}).}
\end{small}
\end{deluxetable}   

%-----------------------------------------------------------------------------%
\section{Analysis}
\label{S3}
%-----------------------------------------------------------------------------%

%-----------------------------------------------------------------------------%
\subsection{The Neutral ISM of UGC\,11411 and UGC\,8245}
\label{S3.1}
%-----------------------------------------------------------------------------%

\begin{figure*}[!ht]
\epsscale{1.2}
\plotone{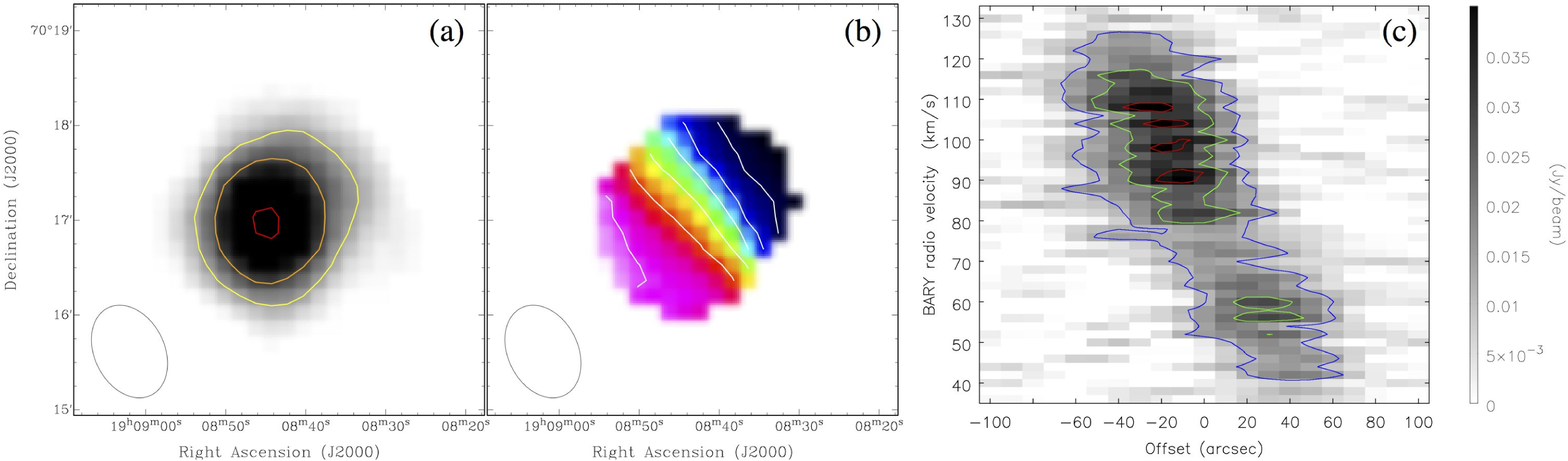}
\epsscale{1.0}
\caption{HI images of UGC\,11411. Panel (a) shows the HI mass surface
  density image, with contours overlaid at the (1.5, 3,
  6)\,$\times$10$^{20}$ cm$^{-2}$ levels, represented by yellow,
  orange, and red, respectively.  Panel (b) shows the intensity
  weighted HI velocity field, with contours overlaid in 10
  \kms\ intervals between 56 \kms\ and 106 \kms.  The beam size is
  shown in the bottom left of panels (a) and (b).  Panel (c) shows a
  major-axis position-velocity slice.  The slice is centered at
  ($\alpha$,$\delta$) = (19$^{\rm h}$08$^{\rm m}$42.5$^{\rm s}$,
  $+$70\arcdeg17$^{\prime}$05.7$^{\prime\prime}$), is 5 pixels
  (50\arcsec) wide, and is at a position angle of
  $+$135\arcdeg\ measured east from north.  The contours are shown at
  the 4\,$\sigma$ (blue), 8\,$\sigma$ (green), and 12\,$\sigma$ (red)
  levels.}
\label{UGC11411.HI} 
\end{figure*}

\begin{figure*}
\epsscale{1.2}
\plotone{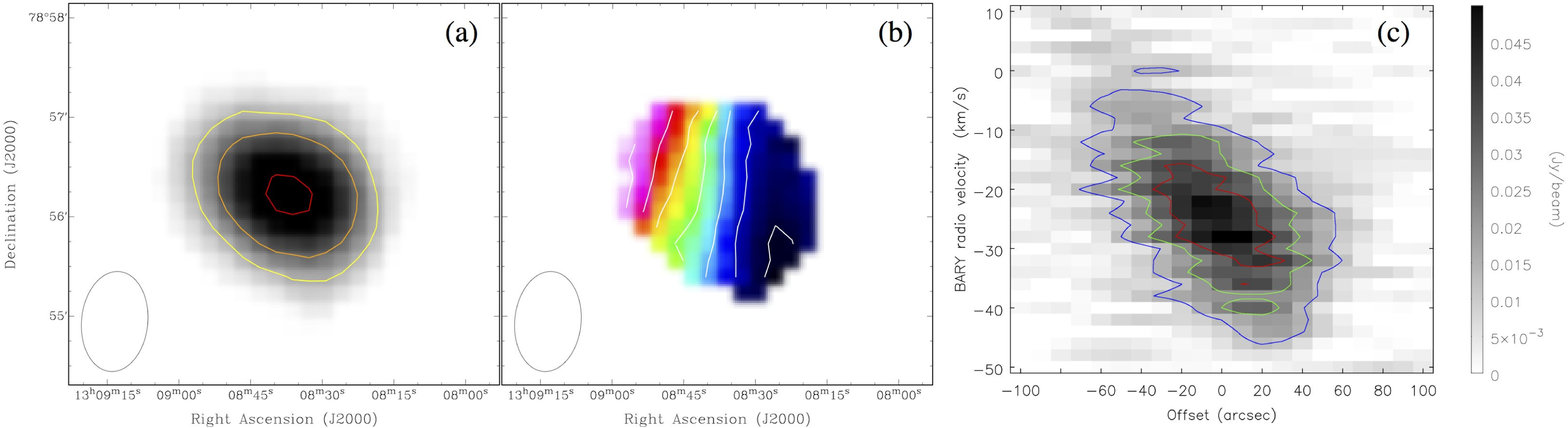}
\epsscale{1.0}
\caption{HI images of UGC\,8245. Panel (a) shows the HI mass surface
  density image, with contours overlaid at the (1.25, 2.5,
  5)\,$\times$10$^{20}$ cm$^{-2}$ levels, represented by yellow,
  orange, and red, respectively.  Panel (b) shows the intensity
  weighted HI velocity field, with contours overlaid in 3
  \kms\ intervals between $-$31 \kms\ and $-$13 \kms. The beam size is
  shown in the bottom left of panels (a) and (b).  Panel (c) shows a
  major-axis position-velocity slice.  The slice is centered at
  ($\alpha$,$\delta$) = (13$^{\rm h}$08$^{\rm m}$35.3$^{\rm s}$,
  $+$78\arcdeg56$^{\prime}$14.3$^{\prime\prime}$), is 5 pixels
  (50\arcsec) wide, and is at a position angle of
  $+$70\arcdeg\ measured east of north.  The contours are shown at the
  4\,$\sigma$ (blue), 8\,$\sigma$ (green), and 12\,$\sigma$ (red)
  levels.}
\label{UGC8245.HI} 
\end{figure*}

HI images of UGC\,11411 and UGC\,8245 are presented in
Figures~\ref{UGC11411.HI} and \ref{UGC8245.HI}, respectively.  While
the angular resolution is coarse, the galaxies are formally resolved,
both along the major and the minor axes of the beams.  The moment zero
maps localize the HI gas and provide high surface brightness
sensitivity; the elongation of each galaxy (especially in the
directions parallel to the synthesized beam minor axis) is physically
meaningful and is not a resolution effect.  The HI column densities
peak above 5\,$\times$\,10$^{20}$ cm$^{-2}$ in each source; higher
angular resolution images are expected to increase the column
densities within the inner disk region of each source.

The moment one maps, representing intensity-weighted velocity, reveal
the bulk projected rotation of each source.  While we anticipate that
higher angular resolution images will reveal significant kinematic
substructure within each galaxy's disk, our low resolution images
reveal coherent projected rotation of each galaxy (see also discussion
in \S~\ref{S3.3}).  The intensity weighted velocity fields shown in
Figures~\ref{UGC11411.HI} and \ref{UGC8245.HI} are characterized by
nearly-parallel isovelocity contours in the inner region of each disk.
Importantly, while the northerly declination of each galaxy results in
an asymmetric beam shape, there is a fortuitous alignment of the beam
minor axis and the kinematic major axis in each galaxy; the signatures
of rotation can be interpreted with confidence.


%-----------------------------------------------------------------------------%
\subsection{Stellar Populations and Star Formation in UGC\,11411 and UGC\,8245}
\label{S3.2}
%-----------------------------------------------------------------------------%

\begin{figure*}[!ht]
\epsscale{1.2}
\plotone{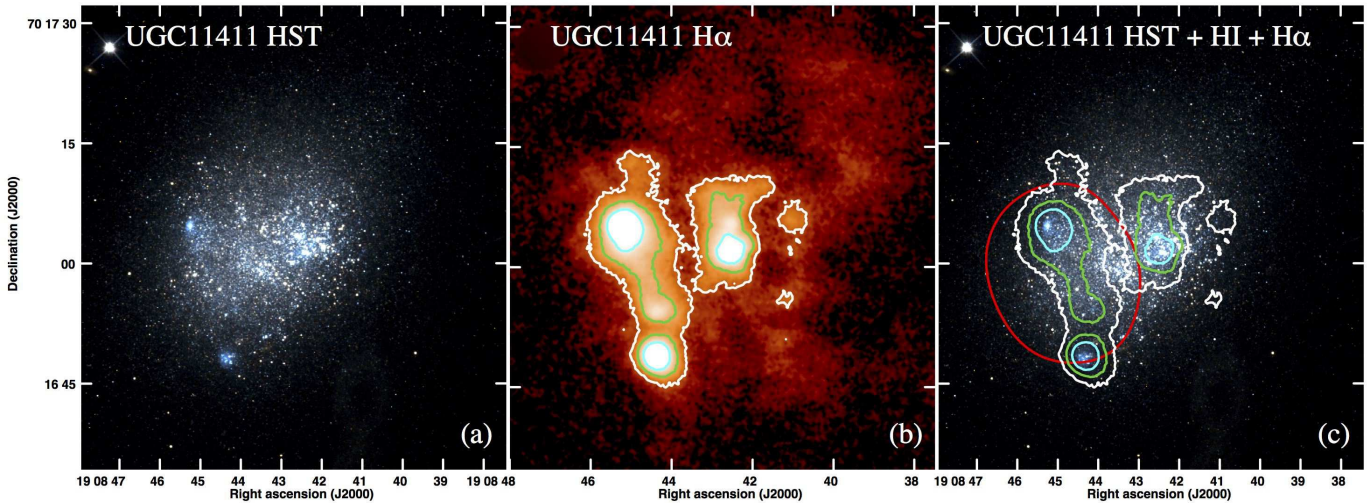}
\epsscale{1.0}
\caption{HST 3-color image (a), SAO telescope continuum-subtracted
  \halpha\ image (b), and the HST 3-color image compared with HI and
  \halpha\ (c) of UGC\,11411. The galaxy harbors widespread, high
  surface brightness \halpha\ emission.  The display of the
  \halpha\ image is logarithmic, with contours separated by 0.4 dex
  each; note the extended \halpha\ emission to the northwest that is
  not contoured but is outside of the optical main body. The red
  contour in panel (c) is the same as shown in
  Figure~\ref{UGC11411.HI} (N$_{\rm HI} $=$ $6\,$\times$\,10$^{20}$
  cm$^{-2}$.)}
% Halpha contours ar 3.1, 3.5, 3.9
\label{UGC11411.optical} 
\end{figure*}

UGC\,11411 is an actively star-forming dwarf galaxy.  As shown in
Figure~\ref{UGC11411.optical}(a), HST observations of UGC\,11411
resolve individual stars; this provides a CMD-based distance
measurement of D $=$ 6.58$\pm$0.12 Mpc (Tully \etal\ 2013).  Nebular
emission is prominent in the HST images, and numerous stellar clusters
and massive star formation complexes are evident.  The
continuum-subtracted H$\alpha$ image shown in
Figure~\ref{UGC11411.optical}(b) reveals widespread, high surface
brightness \halpha\ emission throughout the main body of UGC\,11411.
Further, there is diffuse \halpha\ nebulosity extending well beyond
the stellar disk as traced by HST; note by examining
Figure~\ref{UGC11411.optical}(b) that this extended \halpha\ emission
is especially prominent in the northwest region. The instantaneous
massive star formation rate derived from the \halpha\ image,
log[SFR$_{\rm H\alpha}$/(\msun\,yr$^{-1}$)] = $-$1.84, compares well
with the longer-timescale, UV-based star formation rate of
log[SFR$_{\rm FUV}$/(\msun\,yr$^{-1}$)] = $-$1.72 as found by
\citet{karachentsev13}.  No constraints on the metallicity of the
nebular gas are currently available.

\begin{figure*}
\epsscale{1.2}
\plotone{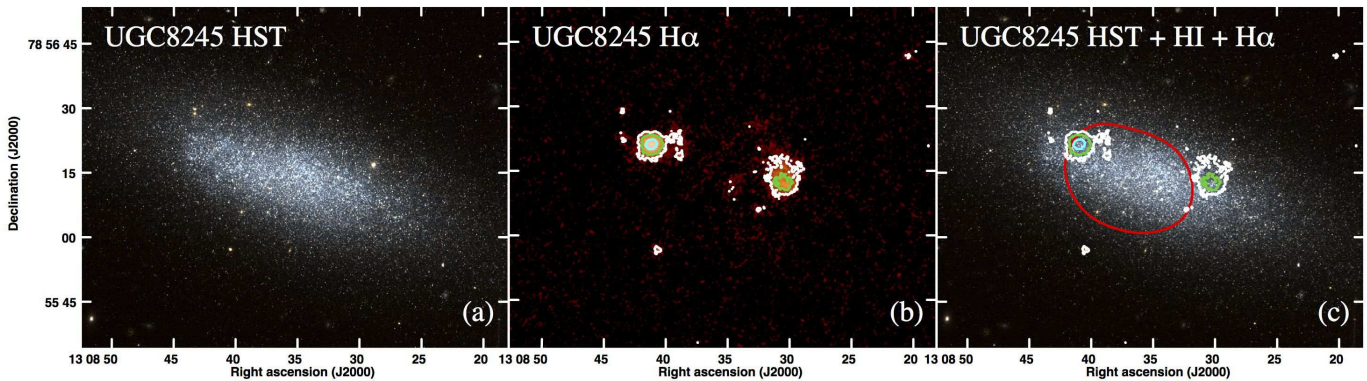}
\epsscale{1.0}
\caption{HST 3-color image (a), SAO telescope continuum-subtracted
  \halpha\ image (b), and the HST 3-color image compared with HI and
  \halpha\ (c) of UGC\,8245. The galaxy hosts only two relatively
  compact and faint HII regions.  The display of the \halpha\ image is
  logarithmic, with contours separated by 0.4 dex each. The red
  contour in panel (c) is the same as shown in
  Figure~\ref{UGC8245.HI} (N$_{\rm HI} $=$ $5\,$\times$\,10$^{20}$
  cm$^{-2}$.)}
% Halpha contours at 2.8, 3.2, 3.6
\label{UGC8245.optical} 
\end{figure*}

UGC\,8245 is a more quiescent system than UGC\,11411.  As shown in
Figure~\ref{UGC8245.optical}, HST observations resolve the stellar
population in detail, producing a CMD-based distance D $=$
4.72$\pm$0.07 Mpc (Tully \etal\ 2013).  While the stellar population
is blue in the inner disk, UGC\,8245 does not show any prominent
nebular emission in the HST image.  The ground-based \halpha\ image
reveals two faint star formation complexes; neither is aligned
spatially with obvious stellar clusters in the HST images.  There is
no diffuse or extra-planar \halpha\ emission detected at this
sensitivity level.  The instantaneous and UV-based star formation
rates are lower in UGC\,8245 than in UGC\,11411: log[SFR$_{\rm
    H\alpha}$/(\msun\,yr$^{-1}$)] = $-$3.02 and log[SFR$_{\rm
    FUV}$/(\msun\,yr$^{-1}$)] = $-$2.46, respectively \citep{KK13}.
Interestingly, the brighter HII region on the eastern end of the disk
was observed spectroscopically by \citet{berg12}, resulting in a
sub-Solar oxygen abundance measurement (Z $\simeq$ 8-12\% Z$_{\odot}$
depending on which strong-line method is applied).

A spatially resolved study of the relationships between neutral HI gas
and ongoing star formation (as was presented in {Teich
  \etal\ 2016}\nocite{teich16}) is not possible at our present angular
resolution.  However, comparisons of the locations of the highest HI
mass surface densities and \halpha\ surface brightnesses are possible;
Figures~\ref{UGC11411.optical}(c) and \ref{UGC8245.optical}(c) show
the highest HI column density contour superposed on contours of the
\halpha\ emission.  In both galaxies, the brightest \halpha\ regions
are either co-spatial with, or very close to, the HI maxima.  The
compact HII regions in UGC\,8245 are confined to the gas-rich inner
disk.  Similarly, in UGC\,11411, the major star formation complexes
are within or very close to the dense HI gas.  However,
Figure~\ref{UGC11411.optical} highlights the remarkable extent of the
\halpha\ emission outside of the inner disk.


We use the HST and ground-based images to determine the inclination of
each galaxy; a cursory examination of Figures~\ref{UGC11411.optical}
and \ref{UGC8245.optical} reveals that UGC\,11411 is more face-on than
UGC\,8245.  Following \citet{karachentsev13}, a first measurement of
the inclination is obtained by measuring the apparent axial ratio of
each galaxy in the broad-band images that were acquired with the
continuum-subtracted \halpha\ images.  Assuming an intrinsic axial
ratio of 0.42, we find that $i$\,$=$\,52\arcdeg\ and
$i$\,$=$\,90\arcdeg\ for UGC\,11411 and UGC\,8245, respectively.  This
method can be compared with a measurement of the axial ratio of each
galaxy in the HST images using the custom software package
\textsc{CleanGalaxy} \citep{hagen2014}.  Here, surface brightness
contours as a function of galactocentric radius are fit to the HST
image after it has been cleaned of foreground and background objects
and smoothed with a spatial Gaussian filter.  Using this method and
assuming the same intrinsic axial ratio of 0.42, we derive
inclinations of $i$\,$=$\,37\arcdeg\ and $i$\,$=$\,75\arcdeg\ for
UGC\,11411 and UGC\,8245, respectively.

For consistency with the results presented in \citet{teich16} and
\citet{cannon16}, we use both values of inclination for each galaxy in
the discussion that follows.  The lower inclinations dervied using the
HST images compared to those using the ground-based images can be
attributed to the different surface brightness sensitivities of the
two sets of images, and also in part to the differences in the methods
applied to excise foreground and background objects.  It is
interesting to note that \citet{roychowdhury10} suggest that the
gaseous disks of faint dwarf irregular galaxies are quite thick, with
a mean intrinsic axial ratio q$_{\rm 0}$ $\simeq$ 0.57.  The HST-based
inclination of UGC\,8245 reaches 90\arcdeg\ when q$_{\rm 0}$\,$>$
0.48; if q$_{\rm 0}$\,$=$ 0.57 then the HST-based inclination of
UGC\,11411 increases to 42\arcdeg.  We note that for UGC\,8245, which
is effectively edge-on, the adopted inclination makes very little
difference in the determination of the rotation velocities in
\S~\ref{S3.3}; the effect is more significant for UGC\,11411.

%-----------------------------------------------------------------------------%
\subsection{Rotational Dynamics of UGC\,11411 and UGC\,8245}
\label{S3.3}
%-----------------------------------------------------------------------------%

Given the small number of synthesized beams subtending each galaxy, we
are unable to fit detailed two or three-dimensional models to derive
their rotational dynamics.  We therefore apply the techniques
presented in \citet{mcnichols16} to create the position-velocity
slices shown in Figures~\ref{UGC11411.HI}(c) and \ref{UGC8245.HI}(c).
The position angle of each slice is determined iteratively by
maximizing the projected velocity along the kinematic major
axis; this position angle is checked by examining the position angle
of the isovelocity contours at the systemic velocity of each galaxy
and then rotating each by $+$90\arcdeg.  The central position of each
slice is at the midpoint of the isovelocity contour at systemic
velocity.  Each position-velocity slice is 5 pixels wide and thus
wider than the HI beam minor axis for each galaxy.

For UGC\,11411, the position-velocity slice is centered at
($\alpha$,$\delta$) = (19$^{\rm h}$08$^{\rm m}$42.5$^{\rm s}$,
$+$70\arcdeg17$^{\prime}$05.7$^{\prime\prime}$), following a position
angle of 135\arcdeg\ measured east of north.  As shown in
Figure~\ref{UGC11411.HI}(c), the projected rotation of the galaxy
shows the characteristic solid-body signature that is typical of
low-mass galaxies (e.g., {McNichols \etal\ 2016}\nocite{mcnichols16}).
We use the 4\,$\sigma$ contours in Figure~\ref{UGC11411.HI} to define
the total projected velocity ($\sim$87 \kms) and angular extent
($\pm$60\arcsec\ from the dynamical center) of the gas.  Correcting
for inclination (see discussion above) and assuming symmetric
projected rotation about the dynamical center at $\pm$43.5 \kms, the
corresponding dynamical mass estimates are M$_{\rm dyn}$ =
2.3$\times$10$^9$ \msun\ ($i$ $=$ 37\arcdeg) and M$_{\rm dyn}$ =
1.4$\times$10$^9$ \msun\ ($i$ $=$ 52\arcdeg).

The major-axis position-velocity slice for UGC\,8245 shows similar
solid-body rotation characteristics as that for UGC\,11411, although
with a smaller projected velocity.  As Figure~\ref{UGC8245.HI} shows,
within the 4\,$\sigma$ contours, the total projected velocity is 
$\sim$44 \kms\ spanning an angular offset of $\pm$60\arcsec.
Interestingly, this major-axis slice shows a significant angular
offset at all velocities, with some low surface brightness gas on the
positive-velocity side of the slice; we interpret these features
further in \S~\ref{S4}.  Correcting for inclination (see discussion
above), and again assuming symmetric rotation about the dynamical
center with a total magnitude of $\pm$22 \kms, the resulting dynamical
mass estimates are M$_{\rm dyn}$ = 1.7$\times$10$^8$ \msun\ ($i$ $=$
75\arcdeg) and M$_{\rm dyn}$ = 1.5$\times$10$^8$ \msun\ ($i$ $=$
90\arcdeg).

It is important to note that in the absence of detailed
three-dimensional modeling, the angular resolution of these data
results in a significant uncertainty on the derived dynamical mass of
each galaxy.  Specifically, while Figures~\ref{UGC11411.HI} and
\ref{UGC8245.HI} show high significance (4\,$\sigma$) HI gas at
angular offsets of $\pm$60\arcsec, this is only marginally larger than
the minor axis dimensions of the synthesized beams; we cannot rule out
that beam smearing has artificially enlarged the angular offset from
the minor axis beam size to the estimated $\pm$60\arcsec\ values.
Since M$_{\rm dyn}$ scales linearly with angular offset, the dynamical
mass estimates should be considered uncertain at the 50\% level;
higher angular resolution imaging of each galaxy is warranted.

%-----------------------------------------------------------------------------%
\section{Discussion}
\label{S4}
%-----------------------------------------------------------------------------%

Our new HI observations reveal that UGC\,8245 is only marginally dark
matter dominated.  Correcting the detected 1.4\,$\times$\,10$^{7}$
\msun\ of HI by a factor of 1.35 for helium and metals, the sum of the
total gas mass and the stellar mass is $\sim$9.2\,$\times$\,10$^{7}$
\msun.  The total dynamical mass is $\sim$2 times larger than the
baryonic mass, regardless of which inclination value is adopted for
the disk.  The physical properties of UGC\,8245 are very similar to
those of the more massive SHIELD galaxies discussed in
\citet{mcnichols16}; these galaxies share similar HI and stellar
masses, as well as similar M$_{\rm dyn}$/M$_{\rm bary}$ ratios.

UGC\,11411, on the other hand, is a more dark matter dominated system
that is more massive than all of the SHIELD galaxies discussed in
\citet{mcnichols16}.  Using the larger value of inclination 
($i$\,$=$\,52\arcdeg), the total baryonic mass (M$_{\rm bary}$ $=$
1.1\,$\times$\,10$^{8}$ \msun) is smaller than the dynamical mass by a
factor of more than 10.  If the inclination is in fact lower ($i$ $=$
37\arcdeg) then the dynamical mass exceeds the baryonic mass by a
factor of 20.  Such a highly dark matter dominated halo is unusual
amongst the population of local volume dwarf galaxies.

Both of the galaxies presented in this work show the characteristic
signatures of solid-body rotational dynamics.  Interestingly, these
two galaxies probe the beginning of the transition regime from
rotationally-dominated to pressure-dominated systems that is discussed
in detail in \citet{mcnichols16}.  In UGC\,11411 the rotational
velocity is much larger than the velocity dispersion at all radii.  In
contrast, the major axis position-velocity slice of UGC\,8245 shows
both a smaller projected rotation but also a wider velocity dispersion
at all slice positions (and radii).  Using the detailed discussion in
\citet{mcnichols16} as a guide, this is most easily interpreted as the
signature of low projected rotation velocity superposed on the
characteristic velocity dispersion of the HI gas.  The low
(inclination corrected) rotation velocity of UGC\,8245 is approaching
(but still above) the empirical limit at which current observations
can no longer differentiate between pressure-dominated and
rotation-dominated systems.

Each of these galaxies has a low (barycentric) recessional velocity,
and yet each is located well outside the Local Group.  The HST-based
distance measurements are accurate at the $\sim$5\% level, which
allows the peculiar velocities to be fixed with an accuracy of
$\sim$20 \kms\ (V$_{\rm pec}$\,$=$\,$-$114 \kms\ and $-$167 \kms\ for
UGC\,11411 and UGC\,8245, respectively).  Each of the galaxies is
relatively isolated, and thus the negative peculiar velocities are not
likely to be caused by virial motions alone.  Further, the negative
peculiar velocities are not unique; in a wide region of sky
($\sim$3000 square degrees) around the considered galaxies there are
more than a dozen galaxies with similarly low velocities (including 
the two massive spirals M\,101 and NGC\,6946).  UGC\,11411 and UGC\,8245 
appear to be associated with a coherent flow of nearby galaxies
situated at high supergalactic latitude.

%-----------------------------------------------------------------------------%
\acknowledgements
%-----------------------------------------------------------------------------%

The authors are grateful to the NRAO for making the ``Observing for
University Classes'' program available to the astronomical and
teaching communities.  We thank Macalester College and the
Metropolitan State University of Denver for support.  I.K. is
supported by the Russian Science Foundation grant No 14-12-00965.  The
authors thank the anonymous referee for a prompt and insightful report
that helped to strengthen this manuscript.

\facility{VLA, HST, GALEX}

%-----------------------------------------------------------------------------%
\bibliographystyle{apj}                                                 

\begin{thebibliography}{}

\bibitem[Begum et al.(2008)]{begum2008} Begum, A., Chengalur, 
J.~N., Karachentsev, I.~D., Sharina, M.~E., 
\& Kaisin, S.~S.\ 2008, \mnras, 386, 1667 

\bibitem[Berg et al.(2012)]{berg12} Berg, D.~A., Skillman, E.~D.,
  Marble, A.~R., et al.\ 2012, \apj, 754, 98

\bibitem[Cannon et al.(2011)]{cannon11} Cannon, J.~M., Giovanelli, R.,
  Haynes, M.~P., et al.\ 2011, \apjl, 739, L22

\bibitem[Cannon et al.(2016)]{cannon16} Cannon, J.~M., McNichols,
  A.~T., Teich, Y.~G., et al.\ 2016, \aj, 152, 202

\bibitem[Cannon \& Van Moorsel(2017)]{cannon17} Cannon, J.~M., \& Van
  Moorsel, G.~A.\ 2017, American Astronomical Society Meeting
  Abstracts, 229, 336.06

\bibitem[Cook et al.(2014)]{cook14} Cook, D.~O., Dale, D.~A., Johnson,
  B.~D., et al.\ 2014, \mnras, 445, 899

\bibitem[Giovanelli et al.(2005)]{giovanelli05} Giovanelli, R.,
  Haynes, M.~P., Kent, B.~R., et al.\ 2005, \aj, 130, 2598

\bibitem[Hagen et al.(2014)]{hagen2014} Hagen, C., Cannon, J.~M., 
Cave, I., et al.\ 2014, American Astronomical Society Meeting Abstracts 
\#223, 223, \#355.16 

\bibitem[Hunter et al.(2012)]{hunter2012} Hunter, D.~A., 
Ficut-Vicas, D., Ashley, T., et al.\ 2012, \aj, 144, 134 

\bibitem[Karachentsev et al.(2013)]{karachentsev13} Karachentsev,
  I.~D., Makarov, D.~I., \& Kaisina, E.~I.\ 2013, \aj, 145, 101

\bibitem[Karachentsev \& Kaisina(2013)]{KK13}
  Karachentsev, I.~D., \& Kaisina, E.~I.\ 2013, \aj, 146, 46

\bibitem[Kirby et al.(2012)]{kirby2012} Kirby, E.~M., Koribalski, 
B., Jerjen, H., \& L{\'o}pez-S{\'a}nchez, {\'A}.\ 2012, \mnras, 420, 2924 

\bibitem[Lelli et al.(2014)]{lelli14} Lelli, F., Verheijen, M., \&
  Fraternali, F.\ 2014, \aap, 566, A71

\bibitem[McNichols et al.(2016)]{mcnichols16} McNichols,
  A.~T., Teich, Y.~G., Nims, E., et al.\ 2016, \apj, 832, 89

\bibitem[Meyer et al.(2004)]{meyer04} Meyer, M.~J., Zwaan, M.~A.,
  Webster, R.~L., et al.\ 2004, \mnras, 350, 1195

\bibitem[Ott et al.(2012)]{ott2012} Ott, J., Stilp, A.~M., 
Warren, S.~R., et al.\ 2012, \aj, 144, 123 

\bibitem[Roychowdhury et al.(2010)]{roychowdhury10} Roychowdhury,
  S., Chengalur, J.~N., Begum, A., \& Karachentsev, I.~D.\ 2010,
  \mnras, 404, L60

\bibitem[Swaters et al.(2002)]{swaters2002} Swaters, R.~A., van
  Albada, T.~S., van der Hulst, J.~M., \& Sancisi, R.\ 2002, \aap,
  390, 829

\bibitem[Teich et al.(2016)]{teich16} Teich, Y.~G., McNichols, A.~T.,
  Nims, E., et al.\ 2016, \apj, 832, 85

\bibitem[Tully et al.(2013)]{tully13} Tully, R.~B., Courtois, H.~M.,
  Dolphin, A.~E., et al.\ 2013, \aj, 146, 86

\bibitem[Wang et al.(2017)]{wang17} Wang, J., Koribalski, B.~S., 
Jarrett, T.~H., et al.\ 2017, \mnras, in press (ArXiV/1708.02744)

\end{thebibliography}

%-----------------------------------------------------------------------------%

%%%%%%%%%%%%%%%%%%%%%%%%%%%%%%%%%%%%%%%%%%%%%%%%%%%%%%%%%%%%%%%%%%%%%%%%%%%%%%%%%%

%%%%%%%%%%%%%%%%%%%%%%%%%%%%%%%%%%%%%%%%%%%%%%%%%%%%%%%%%%%%%%%%%%%%%%%%%%%%%%%%%%

\end{document}